\documentclass{llncs}
\pdfoutput=1

\newif\ifThesis
\Thesisfalse

\newif\ifcomment
\commentfalse

\newif\iflongversion
\longversiontrue

\newif\ifcomment
\commentfalse

\newif\ifACM
\ACMfalse

\newif\ifLNCS
\LNCStrue

\newif\ifdraft
\draftfalse

\newcommand{\paperpath}{./}

\usepackage{lhelp}
\usepackage{amsfonts,latexsym}
\usepackage{amsmath}
\usepackage{amssymb}
\usepackage[font=footnotesize]{caption}
\usepackage{tikz}
\usepackage{listings}
\usetikzlibrary{positioning,calc}
\usepackage{hyperref}
\usepackage{cleveref}
\usepackage{color}
\usepackage{graphics}
\usepackage{subcaption}

\usepackage{booktabs}

\newtheorem{thm}{Theorem}
\newtheorem{prop}{Proposition}
\newtheorem{defn}{Definition}
\newtheorem{lem}{Lemma}
\newtheorem{cor}{Corollary}

\usepackage{amssymb}
\newcommand*{\QEDB}{\hfill\ensuremath{\square}}%

\def\C {\ensuremath{\mathbb{C}}}

\def\Q {\ensuremath{\mathbb{Q}}}
\def\R {\ensuremath{\mathbb{R}}}

\def\KK {\ensuremath{\mathbb{K}}}

\newcommand{\Maple}{\textsc{Maple}}
\newcommand{\bpas}{\textsc{bpas}}

\newcommand{\mpsolve}{\textsc{MPSolve}}

\usepackage{paralist}
        {\begin{enumerate}[#1]}{\end{enumerate}%
         \vspace{-.6\baselineskip}}
        {\begin{compactenum}[#1]}{\end{compactenum}}

\newcommand{\bt}{\begin{thm}}
\newcommand{\et}{\end{thm}}
\newcommand{\bco}{\begin{cor}}
\newcommand{\eco}{\end{cor}}
\newcommand{\bl}{\begin{lem}}
\newcommand{\el}{\end{lem}}
\newcommand{\bp}{\begin{prop}}
\newcommand{\ep}{\end{prop}}
\newcommand{\bdf}{\begin{defn}}
\newcommand{\edf}{\end{defn}}
\newcommand{\bpf}{\begin{proof}}
\newcommand{\epf}{\end{proof}}
\newcommand{\bcor}{\begin{correct}}
\newcommand{\ecor}{\end{correct}}
\newcommand{\bci}{\begin{compactitem}}
\newcommand{\eci}{\end{compactitem}}

\newcommand{\be}{\begin{enumerate}}
\newcommand{\ee}{\end{enumerate}}
\newcommand{\bi}{\begin{itemize}}
\newcommand{\ei}{\end{itemize}}
\newcommand{\bd}{\begin{description}}
\newcommand{\ed}{\end{description}}
\newcommand{\bc}{\begin{center}}
\newcommand{\ec}{\end{center}}
\newcommand{\bq}{\begin{quote}\small}
\newcommand{\eq}{\end{quote}}
\newcommand{\beq}{\begin{equation}}
\newcommand{\eeq}{\end{equation}}
\newcommand{\bea}{\setlength\arraycolsep{2pt}\begin{eqnarray}}
\newcommand{\eea}{\end{eqnarray}\setlength\arraycolsep{6pt}}
\newcommand{\bfig}{\begin{figure}}
\newcommand{\efig}{\end{figure}}
\newcommand{\bcom}{}

\newcommand{\viz}{\emph{viz.}}

\newcommand{\s}{\section}
\newcommand{\subs}{\subsection}

\usepackage{color}
\usepackage{listings}
\usepackage{algpseudocode}
\usepackage{algorithm}
\newcommand{\alnam}[1]{\textsc{#1}}

\newcommand{\vect}[1]{\boldsymbol{#1}}

\newcommand{\citep}[1]{\cite{ #1}}

\usepackage{graphicx}
\usepackage{authblk}

\title{Symbolic-Numeric Integration of Rational Functions}
\author{Robert M Corless$^1$, Robert HC Moir$^1$, Marc Moreno Maza$^1$, Ning Xie$^2$\\ 
{$^1$\footnotesize{}Ontario Research Center for Computer Algebra,\\ University of Western Ontario, Canada}\\
{$^2$\footnotesize{}Huawei Technologies Corporation, Markham, ON}}
\institute{}

\begin{document}

\maketitle

\thispagestyle{empty}
\pagestyle{plain}

\ifThesis
\else
\begin{abstract}
\fi
We consider the problem of symbolic-numeric integration of symbolic functions, focusing on rational functions. Using a hybrid method allows the stable yet efficient computation of symbolic antiderivatives while avoiding issues of ill-conditioning to which numerical methods are susceptible. We propose two alternative methods for exact input that compute the rational part of the integral using Hermite reduction and then compute the transcendental part two different ways using a combination of exact integration and efficient numerical computation of roots. The symbolic computation is done within {\bpas}, or Basic Polynomial Algebra Subprograms, which is a highly optimized environment for polynomial computation on parallel architectures, while the numerical computation is done using the highly optimized multiprecision rootfinding package \mpsolve. We show that both methods are forward and backward stable in a structured sense and away from singularities tolerance proportionality is achieved by adjusting the precision of the rootfinding tasks.
\ifThesis
\else
\end{abstract}
\fi

\nocite{bronstein1997symbolic}

\s{Introduction}

\begin{figure}[t]
\centering
\ifLNCS
\includegraphics[scale=0.2]{\paperpath/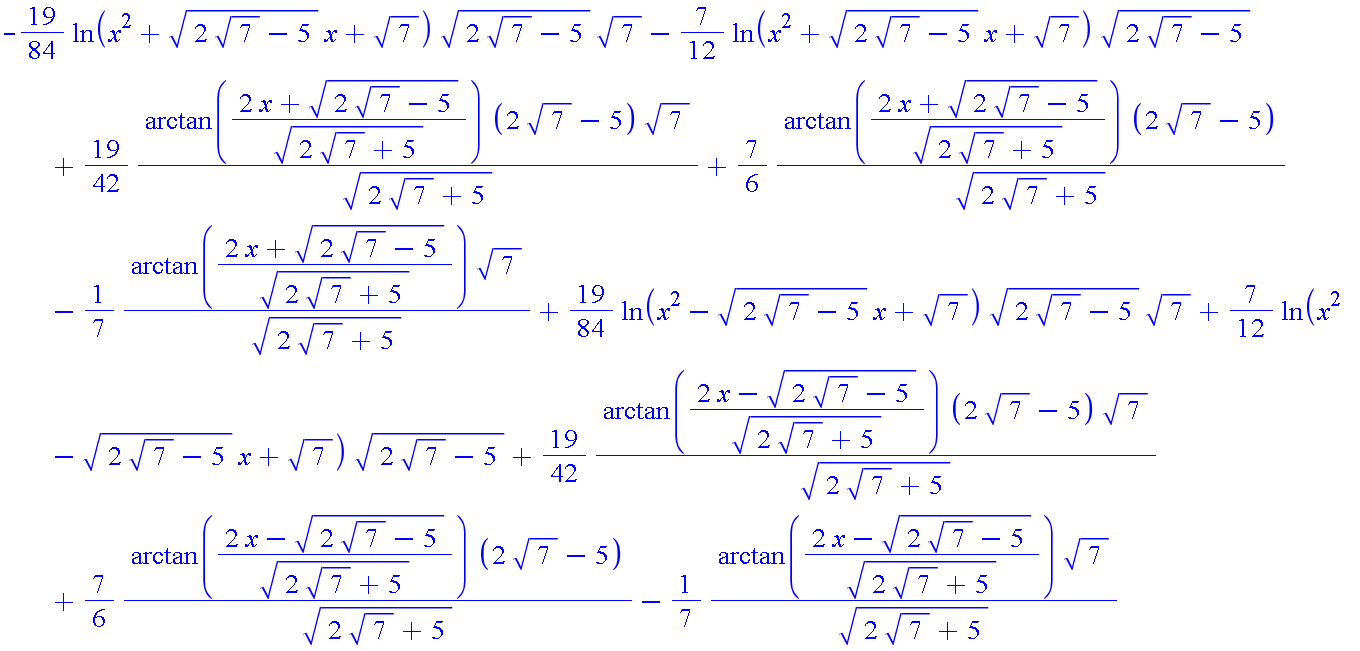}
\fi
\caption{{\Maple} output for the integral $\int\frac{x^2-1}{x^4+5x^2+7}dx$.}
\label{maple-integral}
\end{figure}

Hybrid symbolic-numeric integration of rational functions is
interesting for several reasons. First, a
formula, not a number or a computer program or subroutine, may be
desired, perhaps for further analysis such as by taking
asymptotics. In this case one typically wants an exact
symbolic answer, and for rational functions this is in principle
always possible. However, an exact symbolic answer may be too
cluttered with algebraic numbers or lengthy rational numbers to be
intelligible or easily analyzed by further symbolic
manipulation. See, \emph{e.g.}, Figure \ref{maple-integral}. Discussing symbolic integration,
Kahan \cite{kahan1980handheld} in his typically dry way gives an
example ``atypically modest, out of consideration for the
typesetter'', and elsewhere has rhetorically wondered: ``Have you ever
used a computer algebra system, and then said to yourself as
screensful of answer went by, ``I wish I hadn't asked.''\,'' Fateman
has addressed rational integration \cite{fateman2008revisiting}, as
have Noda and Miyahiro \cite{noda1990symbolic,noda1992hybrid}, for
this and other reasons.

Second, there is interest due to the potential to carry symbolic-numeric
methods for rational functions forward to transcendental integration, since
the rational function algorithm is at the core of more advanced algorithms
for symbolic integration. Particularly in the context of \emph{exact} input, 
which we assume, it can be desirable to have an
intelligible approximate expression for an integral while retaining the
exact expression of the integral for subsequent symbolic computation.
The ability to do this is a feature of one of our algorithms that alternative
approaches, particularly those based on partial fraction decomposition,
do not share.

Besides intelligibility and retention of exact results, 
one might be concerned with numerical
stability, or perhaps efficiency of evaluation. We consider stability
issues in Sections \ref{algorithm-analysis} and \ref{experiments}. 
We remark that the algorithm we present
here has quite superior numerical stability in many cases, and has good 
structured backward error and highly accurate
answers, while providing the more intelligible answers we desire.

We emphasize that the goal of this algorithm is not to produce
numerical values of definite integrals of rational functions, although
it can be used for such. The goal is to produce an intelligible
formula for the antiderivative which is correct in an approximate sense: the
derivative of the answer produced will be another rational function
near to the input, and, importantly, of the same form in that the
denominator will have the correct degrees of its factors in its
squarefree factorization and the residues in its partial fraction
decomposition will also have the same multiplicity.\footnote{Note that strict preservation
of the form of the integrand  is not quite achieved for the PFD method described below, since the 
derivative cannot be simplified into this form without using approximate gcd. Thus, with exact computation, the degree of the numerator and denominator is larger in general than the exact integrand.}

\subs{Symbolic-Numeric integration of Rational Functions}

As indicated above, the combination of symbolic and numerical methods in the integration of rational functions is not new. Noda and Miyahiro \cite{noda1990symbolic,noda1992hybrid} developed a symbolic-numeric, or \emph{hybrid}, method to integrate rational functions based on the use of the approximate symbolic algorithms for noisy data, numerical rootfinding and exact symbolic integration methods. Fateman \cite{fateman2008revisiting} advocates a simpler hybrid approach, largely to produce a fast method that makes symbolic results more useful and more palatable, avoiding the ``surd'' or ``RootOf'' notation in the results of symbolic integrations. Both approaches work with the assumption that the input rational function has floating point coefficients.

For the extant symbolic-numeric algorithms for rational function integration, the approach is to be as sparing as possible in the use of symbolic algorithms to minimize their expense, in particular given that floating point input is assumed to be imprecise. In contrast, given that our working assumption is that the input rational function is \emph{exact}, the present paper is dealing with a somewhat different problem, \viz, the approach involves the injection of numerical methods into an otherwise purely symbolic context. As was pointed out above, the reasons such an approach is desirable include intelligibility, retention of exact results and stable or efficient evaluation.Since it is accuracy, speed and stability that matter in the context of scientific computing, a symbolic package that provides a suitable balance of these desiderata in a way that can be merged seamlessly with other scientific computations, as our implementation provides, has considerable advantages over CAS style symbolic computation with exact roots.

The usual approach to symbolic integration here begins with a rational function $f(x)=A(x)/B(x)\in\mathbb{Q}(x)$, with $\deg(A)<\deg(B)$ (ignoring any polynomial part, which can be integrated trivially) and computes an integral in two stages:
\bci[\quad\ \scriptsize$\bullet$]
\item rational part: computes a rational function $C/D$ such that
\beq
\label{rat-part}
\int\hspace{-3pt}f(x)\,dx=\frac{C(x)}{D(x)}+\int\frac{G(x)}{H(x)}dx,
\eeq
where the integral on the right hand side evaluates to a transcendental function (log and arctan terms);
\item transcendental part: computes the second (integral) term of the expression \eqref{rat-part} above yielding, after post-processing,
\beq
\label{tran-part}
\int\hspace{-4pt}\,f(x)dx=\frac{C(x)}{D(x)}+\sum v_i\log(V_i(x))+\sum w_j\,\text{arctan}(W_j(x)),
\eeq
$V_i,W_j\in\mathbb{K}[x]$, with $\mathbb{K}$ being some algebraic extension of $\mathbb{Q}$.
\eci

In symbolic-numeric algorithms for this process some steps are replaced by numeric or quasi-symbolic methods. Noda and Miyahiro use an approximate Horowitz method (involving  approximate squarefree factorization) to compute the rational part and either the Rothstein-Trager (RT)\label{symbol:rt} algorithm or (linear/quadratic) partial fraction decomposition (PFD)\label{symbol:pfd} for the transcendental part (see Section~\ref{review} for a brief review of these algorithms). The algorithm considered by Fateman avoids the two stage process and proceeds by numerical rootfinding of the denominator $B(x)$ (with multiplicities) and PFD to compute both rational and transcendental parts. In both cases, the working assumption is that the input uses floating point numbers that are subject to uncertainty or noise and 
the numerical algorithms use double precision.

Part of the power of symbolic algorithms is their ability to preserve structural features of a problem that may be very difficult to preserve numerically. Particularly given our focus on exact input, then, we are interested in preserving as much structure of the problem as possible if we are to resort to the use of numerical methods for rootfinding. Our implementation relies on the sophisticated rootfinding package \mpsolve, which provides \emph{a posteriori} guaranteed bounds on the relative error of all the roots for a user-specified tolerance $\varepsilon$. To balance efficiency of computation and structure-preservation, we use more efficient symbolic algorithms where possible, such as the Hermite method for the rational part, and consider two methods of computing the transcendental part, one that computes the exact integral using the Lazard-Rioboo-Trager (LRT)\label{symbol:lrt} method followed by numerical approximation, and the other that uses a multiprecision PFD method to compute a nearby integrand that splits over $\mathbb{Q}$ and then performs a structured integration. For more details on the symbolic algorithms, see Section~\ref{review}. Our symbolic-numeric algorithms are discussed in Section~\ref{algorithm}. 

The advantage of combining multiprecision numerical software with symbolic algorithms is that it allows for the user to specify a tolerance on the error of the symbolic-numeric computation. This, together with \emph{structured} backward and forward error analysis of the algorithm, then allows the result to be interpreted in a manner familiar to users of numerical software but with additional guarantees on structure-preservation. We provide such an analysis of the structured error of our algorithm in Section \ref{algorithm-analysis}.

An interesting feature of the backward stability of the algorithm is that it follows that the computed integral can be regarded as the \emph{exact} integral of a slightly perturbed input integral, and, as stated previously, of the correct form (modulo the need for approximate gcd). Insofar as the input rational function is the result of model construction or an application of approximation theory it is subject to error. Thus, the input, though formally exact, is nevertheless still an approximation of a system it represents. Assuming the model approximation error is small, this means that the ``true'' rational function the input represents is some nearby rational function in a small neighbourhood of $f(x)$, for example in the sense of the space determined by variation of the coefficients of $f$. The backward stability therefore shows that the integral actually computed is also a nearby rational function within another small neighbourhood, for which we have some control over its size. In a manner similar to numerical analysis, then, by an appropriate choice of tolerance, we can ensure that the latter neighbourhood is smaller than the former, so that the numerical perturbation of the problem is smaller than the model approximation error. The upshot of this is that the use of backward error analysis shows how a symbolic-numeric algorithm can be compatible with the \emph{spirit} of uncertain input, even if the input is assumed to be exact. That is, we are assuming that the modeling procedure got the ``right kind'' of rational function to give as input, even if its data is uncertain in other ways.

This shows how a backward error analysis can be useful even in the context of exact integration. In the general case of exact input, however, a backward error analysis alone is not enough. This is why we also provide a forward error analysis, to provide \emph{a posteriori} assurance of a small forward error sufficiently far away from singularities in the integrand. We also provide such an analysis in Section  \ref{algorithm-analysis}. 

Our algorithm may be adapted to take truly uncertain input by using additional symbolic-numeric routines, such as approximate GCD and approximate squarefree factorization, in order to detect nearby problems with increased structure. Such an approach would shift the problem onto the pejorative manifold\label{symbol:pejman} (nearest most singular problem, see \cite{kahan1972conserving}), which protects against ill-conditioning of the problem. The symbolic-numeric structured PFD integration we propose already takes this approach. Moreover, the structured error analysis of our algorithm then entails that the problem stays on the pejorative manifold after the computation of the integral. Since there have been considerable advances in algorithms for approximate polynomial algebra since the time of writing of \cite{noda1992hybrid}, such as the \textsc{ApaTools}\label{symbol:apatools} package of Zeng \cite{zeng2008apatools}, the combination of error control and singular problem detection could yield a considerable advance over the early approach of Noda and Miyahiro.
\section{Methods for Exact Integration of Rational Functions}
\label{review}

We begin by reviewing symbolic methods for integrating rational functions.\footnote{The following review is based in part on the ISSAC 1998 tutorial~\cite{bronstein1998symbolic} 
and the landmark text book~\cite{bronstein1997symbolic} of M.~Bronstein.}
Let $f \in {\R}(x)$ be a rational function over {\R} not belonging to ${\R}[x]$.  There exist polynomials $P, A, B \in {\R}[x]$ such that we have $f = P + A/B$ with ${\gcd}(A,B) = 1$ and ${\deg}(A) < {\deg}(B)$. Since $P$ is integrated trivially, we ignore the general case and assume that $f=A/B$ with ${\deg}(A) < {\deg}(B)$. Furthermore, thanks to Hermite reduction, one can extract the rational part of the integral, leaving a rational function $G/H$, with ${\deg}(G) < {\deg}(H)$ and $H$ squarefree, remaining to integrate. For the remainder of this section, then, we will assume that the function to integrate is given in the form $G/H$, with ${\deg}(G) < {\deg}(H)$ and $H$ squarefree.

\ifThesis
\smallskip\noindent{\sf Partial-fraction decomposition (PFD) 
\else
\smallskip\noindent{\small \bf Partial-fraction decomposition (PFD) 
\fi
algorithm.}
The partial fraction decomposition algorithm for rational functions in $\mathbb{R}(x)$ can be presented in different ways, depending on whether one admits complex numbers in expressions. We present a method based upon a complete factorization of the denominator over $\mathbb{C}$, followed by its conversion into an expression containing only constants from~$\mathbb{R}$.

Consider the splitting of $H$ expressed in the form
$$H = p \prod_{i=1}^{n} \, (x - \alpha_i)  \ 
                \prod_{j=n+1}^{n+m} \, \left[(x - (\alpha_j + i\,\beta_j))(x - (\alpha_j - i\,\beta_j))\right],$$
separating real roots from complex conjugate pairs, where $p,\alpha_k, \beta_k \in {\R}$. Then there exist $a_k$ and $b_k$ such that
\begin{equation}
\label{pfd}
\frac{G}{H} = \sum_{i=1}^{n} \frac{a_i}{x - \alpha_i}
               +  \sum_{j=n+1}^{n+m}  \left[\frac{a_j + i\,b_j}{(x - (\alpha_j + i\,\beta_j))}+\frac{a_j - i\,b_j}{(x - (\alpha_j - i\,\beta_j))}\right].
\end{equation}
The numerator quantities $c_k=a_k+ i\,b_k$ corresponding to the roots $\gamma_k=\alpha_k+i\,\beta_k$ we call \emph{residues} by analogy to complex analysis. Note that in the case here where $H$ is squarefree, the residues can be computed by the formula $c_k=c(\gamma_k)=G(\gamma_k)/H'(\gamma_k)$.

The real root terms are easily integrated to yield terms of the form $a_i \, {\log}(x - \alpha_i)$. Extracting terms of the form 
\ifThesis
$$a_j\left[(x - (\alpha_j + i\,\beta_j))^{-1}+(x - (\alpha_j - i\,\beta_j))^{-1}\right]$$
\else
$a_j\left[(x - (\alpha_j + i\,\beta_j))^{-1}+(x - (\alpha_j - i\,\beta_j))^{-1}\right]$
\fi
 from \eqref{pfd} we obtain pairs of complex log terms that can be combined to form a single real log term of the form $a_j\, {\log}(x^2 - 2 \alpha_j x + \alpha_j^2 + \beta_j^2)$. 
Extracting terms of the form 
\ifThesis
$$i\,b_j\left[(x - (\alpha_j + i\,\beta_j))^{-1}-(x - (\alpha_j - i\,\beta_j))^{-1}\right]$$
\else
$i\,b_j\left[(x - (\alpha_j + i\,\beta_j))^{-1}-(x - (\alpha_j - i\,\beta_j))^{-1}\right]$
\fi
 from \eqref{pfd} and making use of the observation of Rioboo that $\frac{d}{dx}i\log\left(\frac{X+iY}{X-iY}\right)=\frac{d}{dx}2\,\text{arctan}(X/Y)$, for $X,Y\in\mathbb{R}[x]$ (see \cite{bronstein1997symbolic}, pp. 59\emph{ff.}), we obtain a term in the integral of the form  $2b_{j}\, {\text{arctan}}\left(\frac{\alpha_j-x}{\beta_j}\right)$.

Where there are repeated residues in the PFD it is possible to combine terms of the integral together.  The combination of logarithms with common $a_k$ simply requires computing the product of their arguments. For the arctangent terms the combination of terms with common $b_k$ can be accomplished by recursive application of the rule
\begin{equation}
\text{arctan}\left(\frac{X}{Y}\right) + \text{arctan}\left(\frac{\alpha-x}{\beta}\right) \rightarrow \text{arctan}\left(\frac{X(\alpha-x) - \beta Y}{Y(\alpha-x)+\beta X}\right),
\end{equation}
which is based on the fact that $\log(X+i\,Y) + \log((\alpha-x)+i\,\beta) = \log((X(\alpha-x) - \beta Y) + i\,(Y(\alpha-x)+\beta X))$ and Rioboo's observation noted above.

A major computational bottleneck of the symbolic algorithms based on a PFD is the necessity of factoring polynomials into irreducibles over {\R} or {\C} (and not just over {\Q}) thereby introducing algebraic
numbers even if the integrand and its integral are both in ${\Q}(x)$.
Unfortunately, introducing algebraic numbers may be necessary: any field containing an integral of $1 / (x^2 + 2)$ contains $\sqrt{2}$ as well.
A result of modern research are so-called {\em rational} algorithms that compute as much of the integral as can be kept within ${\Q}(x)$, and compute the minimal algebraic extension of ${\KK}$ necessary to express the integral.

\ifThesis
\smallskip\noindent{\sf The Rothstein-Trager theorem.}
\else
\smallskip\noindent{\small \bf The Rothstein-Trager theorem.}
\fi

It follows from the PFD of $G/H$, \emph{i.e.}, $G/H=\mbox{$\sum_{i=1}^nc_i/(x-\gamma_i)$}$, $c_i,  \gamma_i\in\mathbb{C}$, that
\begin{equation}
\int \frac{G}{H}\,dx \ \ = \ \ \sum_{i=1}^{\text{deg}(H)} \, c_i \, {\log}(x - {\gamma}_i)
\end{equation}
where the ${\gamma}_i$ are the zeros of $H$ in {\C} and the $c_i$ are the residues of $G/H$ at the  ${\gamma}_i$. Computing those residues without splitting $H$ into irreducible factors is achieved by
the Rothstein-Trager theorem, as follows.
Since we seek roots of $H$ and their corresponding residues given by evaluating $c=G/H'$ at the roots, it follows that the $c_i$ are exactly the zeros of the \emph{Rothstein-Trager resultant}\label{symbol:rtres} $R \ := \ {\rm resultant}_x(H, G - c H'),$ where $c$ here is an indeterminate. Moreover, the splitting field of $R$ over {\Q} is the minimal algebraic extension of {\Q} necessary to express $\int f $ in the form given by Liouville's theorem, \emph{i.e.}, as a sum of logarithms, and we have
\begin{equation}
\int \frac{G}{H}\,dx \ \ = \ \ \sum_{i=1}^{m} \ \sum_{c \mid U_i(c) = 0} c \ {\log}( {\gcd}(H, G - c H'))
\end{equation}
where $R  \ = \ \prod_{i=1}^{i=m} \, U_i^{e_i}$ is the irreducible factorization of $R$ over {\Q}.

\ifThesis
\smallskip\noindent{\sf The Lazard-Rioboo-Trager algorithm.}
\else
\smallskip\noindent{\small \bf The Lazard-Rioboo-Trager algorithm.}
\fi
Consider the subresultant pseudo-rem\-ain\-der sequence $R_i$, where $R_0=R$ is the resultant  (see p. 115 in~\cite{BL82}) of $H$ and $G - c H'$ w.r.t. $x$. Observe that the resultant $R$ is a polynomial in $c$ of degree ${\deg}(H)$, the roots of which are the residues of $G/H$. Let $U_1 U_2^2 \cdots U_m^m$ be a square-free factorization of $R$. Then, we have 
\begin{equation}
\label{eq:LRT}
\int \frac{G}{H}\,dx \ \ = \ \ \sum_{i=1}^{m} \ \sum_{c \mid U_i(c) = 0} \
                           c \ {\log}( {\gcd}(H, G - c H')),
\end{equation}
which groups together terms of the PFD with common residue, as determined by the multiplicity of $U_i$ in the squarefree factorization. We compute the
\ifThesis
 inner sum 
\else
 sum $\sum_{c \mid U_i(c) = 0} c \ {\log}( {\gcd}(H, G - c H'))$
\fi
 as follows.
If all residues of $H$ are equal, there is a single nontrivial squarefree factor with $i = {\deg}(H)$ yielding $\sum_{c \mid U_i(c) = 0} c \ {\log}(H)$, otherwise, that is, if $i < {\deg}(H)$, the sum is $\sum_{c \mid U_i(c) = 0} c \ {\log}(S_i)$, where $S_i={\rm pp}_x(R_k)$, where $\text{deg}_x(R_k)=i$
and ${\rm pp}_x$ stands for primitive part w.r.t. $x$.
Consequently, this approach requires isolating only the complex roots of the square-free factors 
$U_1, U_2, \ldots, U_m$, whereas methods based on the PFD requires isolating the real or complex roots of the polynomial $H$, where $\text{deg}(H)\geq\sum_i\text{deg}(U_i)$. However, the coefficients of $R$ (and possibly those of $U_1, U_2, \ldots, U_m$) are likely to be larger than those of $H$. Overall, depending on the example, the computational cost of root isolation may put at advantage any of those approaches in comparison to the others.

\s{The Algorithms}
\label{algorithm}

We consider two symbolic-numeric algorithms, both based on Hermite reduction for the rational part and using two distinct methods for the transcendental part, one based on partial fraction decomposition and
the other the Lazard-Rioboo-Trager algorithm, both reviewed in Section~\ref{review}. Following the notation used in equation \eqref{rat-part}, we assume the rational part $C/D$ has been computed and we consider how the transcendental part is computed by the two methods. Both algorithms use {\mpsolve} to control the precision on the root isolation step.

Following the notations used in equation (\ref{eq:LRT}), the LRT-based method proceeds by computing the sub-resultant chain $(R_0, R_1, \ldots)$ and deciding how to evaluate each sum 
\ifLNCS
$\sum_{c \mid U_i(c) = 0} \, c \, {\log}(R_k), \text{deg}(R_k)=i,$ 
\fi
by applying the strategy of Lazard, Rioboo and Trager. However, we compute the complex roots of the polynomials $U_1, U_2, \ldots, U_m$ numerically instead of representing them symbolically as in~\cite{DBLP:journals/jsc/Rioboo03,DBLP:conf/issac/Rioboo92}. Then, we evaluate each sum $\sum_{c \mid U_i(c) = 0} \, c \, {\log}(R_k)$ by an algorithm adapted to this numerical representation of the roots. This method is presented as Algorithm~\ref{symbolicNumericIntegrateLRT}.

The PFD-based method begins by computing numerically the roots $\gamma_i$ of the denominator $H(x)$ and then computes exactly the resulting residues $c_i=c(\gamma_i)=G(\gamma_i)/H'(\gamma_i)$. The numerical rootfinding can break the structure of repeated residues, which we restore by detecting residues that differ by less than $\varepsilon$, the user-supplied tolerance. The resulting partial fraction decomposition can then be integrated using the structure-preserving strategy presented in section \ref{review} above. This strategy allows to algorithm to replicate the structure of the final output from the LRT algorithm as a sum of real logarithms and arctangents. This method is presented as Algorithm~\ref{symbolicNumericIntegratePFD}.

We remark that there can be an issue here in principle as a result of roots of $H$ that are closer than $\varepsilon$. Given the properties of {\mpsolve}, however, this is not an issue in practice, given the ability to compute residues exactly or with sufficiently high precision, because {\mpsolve} isolates roots within regions where Newton's method converges quadratically. In the unlikely event of residues that are distinct but within $\varepsilon$ of each other, the algorithm still results in a small error and is advantageous in terms of numerical stability. This is because identifying nearby roots shifts the problem onto the nearest most singular problem, the space of which Kahan \cite{kahan1972conserving} calls the pejorative manifold, which protects against ill-conditioning.

Both methods take as input a univariate rational function $f(x)=A(x)/B(x)$ over {\Q} with $\text{deg}(B)>\text{deg}(A)$, and a tolerance $\varepsilon>0$.\label{symbol:tol} Both $A(x)$ and $B(x)$ are expressed in the monomial basis. They yield as output an expression
\begin{equation}
\int \hat{f}\,dx=\frac{C}{D}+\sum v_i\log(V_i)+\sum w_j\,\text{arctan}(W_j),
\end{equation}
$V_i,W_j\in\mathbb{Q}[x]$, along with a linear estimate of the forward and backward error. The backward error on an interval $[a,b]$ is measured in terms of $\|\delta(x)\|_{\infty}=\max_{a\leq x\leq b} |\delta(x)|$,\label{symbol:infnorm} where $\delta(x)=f(x)-\frac{d}{dx}\int\hat{f}(x)\,dx$, and the forward error on $[a,b]$ is measured in terms of $\left\|\int(f-\hat{f})\,dx\right\|_{\infty}=\left\|\int \delta(x)dx\right\|_{\infty}$, where $f$ and $f'$ are assumed to have the same constant of integration. Where $f$ has no real singularities, we provide bounds over $\mathbb{R}$, and where $f$ has real singularities the bounds will be used to determine how close to the singularity the error exceeds the tolerance.

The main steps of Algorithm~\ref{symbolicNumericIntegrateLRT} 
and Algorithm~\ref{symbolicNumericIntegratePFD} 
are listed below, where the numbers
between parentheses refer to lines of the pseudo-code below. Both algorithms begin with:
\begin{enumerateshort}
\item[(1-4:)] decompose $\int f\,dx$ into $\frac{C}{D}$ (rational part)
and $\int\hspace{-2pt}\frac{G}{H}\,dx$ (transcendental part) using Hermite reduction;
\end{enumerateshort}
Algorithm~\ref{symbolicNumericIntegrateLRT} then proceeds with:
\begin{enumerateshort}
\item[(5-6:)] compute symbolically the transcendental part $\int\hspace{-2pt}\frac{G}{H}\,dx=\sum_i\sum_{c \mid U_i(c)=0}c\,\cdot\log(S_i(t,x))$
using Lazard-Rioboo-Trager algorithm; in the pseudo-code $\vect{U}$
 is a vector holding the square-free factors of the resultant
while $\vect{S}$ holds the primitive part of elements of the sub-resultant pseudo-remainder sequence corresponding to elements of $\vect{U}$, \viz, such that corresponding to $U_i$ is $S_i=\text{pp}_x(R_k)$, where $\text{deg}_x(R_k)=i$; 
\item[(7:)]  compute the roots $c_k$ of $U_i(c)$ numerically using \textsc{MPSolve} 
to precision $\varepsilon$.
\item[(8-9:)] symbolic post-processing in \textsc{bpas}: computation of the log and arctan terms.
\end{enumerateshort}
After Hermite reduction, Algorithm~\ref{symbolicNumericIntegratePFD} continues with:
\begin{enumerateshort}
\item[(5-6:)]  compute the roots $\gamma_k$ of $H(x)$ numerically using \textsc{MPSolve} 
to precision $\varepsilon$.
\item[(7:)] compute the residues $c_k$ of $G(x)/H(x)$ corresponding to the
approximate roots of $H(x)$ and detect their identity within $\varepsilon$.
\item[(8:)] compute identical residues within $\varepsilon$ and then compute a correspondence $\varphi$ (one-many relation) between a representative residue and its corresponding roots. $\varphi$ correlates indices of selected elements of $\vect{c}$ and indices of elements of $\vect{\gamma}$.
\item[(9-10:)] compute symbolically the transcendental part $\int\hspace{-2pt}\frac{\hat{G}}{\hat{H}}\,dx=\sum v_i\log(V_i)+$\\ $\sum w_j\,\text{arctan}(W_j)$
from the PFD of $\hat{G}(x)/\hat{H}(x)$.
\end{enumerateshort}

Both algorithms complete the integration by processing the arctangent terms, which can be written as $\text{arctan}\left(\frac{X}{Y}\right)$ or $\text{arctan}(X,Y)$, for polynomials $X$ and $Y$, after the integration is complete, using Rioboo's method (described in \cite{bronstein1997symbolic}) to remove spurious singularities. The result is the conversion of the arctangent of a rational function or two-argument arctangent into a sum of arctangents of polynomials.

\vspace{-12pt}

\begin{algorithm}[H]
\scriptsize
\caption{
\scriptsize\alnam{symbolicNumericIntegrateLRT}($f$,$\varepsilon$)\newline $f\in\mathbb{Q}(x)$, $\varepsilon>0$}\label{symbolicNumericIntegrateLRT}
\algnotext{EndIf}
\begin{algorithmic}[1]
\State$(g,h)\gets\alnam{hermiteReduce}(\text{num}(f),\text{den}(f))$ // Note: $g,h\in\mathbb{Q}(x)$
\State$(Quo,Rem)\gets\alnam{euclideanDivide}(\text{num}(h),\text{den}(h))$ // Note: $Quo, Rem \in\mathbb{Q}[x]$
\If{$Quo\neq 0$}
	\State $P \gets \alnam{integrate}(Quo)$
\EndIf
\If{$Rem \neq 0$}
	\State $(\vect{U},\vect{S})\,\gets\,\alnam{integrateLogPart}(Rem,\text{den}(h))$\,// Note: $\vect{U} = (U_i, 1 \leq i \leq m)$ and $\vect{S} = (S_i)$ are vectors with 
coefficients  in $\mathbb{Q}[t]$ and $\mathbb{Q}[t,x]$
respectively
\EndIf
\State $\vect{c} \gets \alnam{rootsMP}(\vect{U},\epsilon)$ 
// Note: $\vect{c} = (c_k)$ are the roots of $U_i$, as  returned by \mpsolve
\State $(\vect{L},\vect{A2}) \gets \alnam{logToReal}(\vect{c},\vect{S})$ // Note: $\vect{L}$ and $\vect{A2}$ are, respectively, vectors of logs and two-argument arctangent terms
	\State $\vect{A} \gets \alnam{atan2ToAtan}(\vect{A2})$ 
\State\Return $(P,g,\vect{L},\vect{A})$
\end{algorithmic}
\end{algorithm}

\vspace{-24pt}

\begin{algorithm}[H]
\scriptsize
\caption{
\scriptsize\alnam{symbolicNumericIntegratePFD}($f$,$\varepsilon$)\newline $f\in\mathbb{Q}(x)$, $\varepsilon>0$}\label{symbolicNumericIntegratePFD}
\algnotext{EndIf}
\begin{algorithmic}[1]
\State$(g,h)\gets\alnam{hermiteReduce}(\text{num}(f),\text{den}(f))$ // Note: $g,h\in\mathbb{Q}(x)$
\State$(Quo,Rem)\gets\alnam{euclideanDivide}(\text{num}(h),\text{den}(h))$ // Note: $Quo, Rem \in\mathbb{Q}[x]$
\If{$Quo\neq 0$}
	\State $P \gets \alnam{integrate}(Quo)$
\EndIf
\If{$Rem \neq 0$}
	\State $\vect{\gamma} \gets \alnam{rootsMP}(\text{den}(h),\epsilon)$ 
	// Note: $\vect{\gamma} = (\gamma_k)$ are the roots of $\text{den}(h)$, as returned by \mpsolve
	\State $\vect{c} \gets \alnam{residues}(Rem,\text{den}(h),\vect{\gamma})$
	// Note: $\vect{c} = (c_k)$ are the residues corresponding to the $\gamma_i$
	\State $\varphi \gets \alnam{residueRootCorrespondence}(\vect{c},\vect{\gamma},\varepsilon)$
	// Note: $\varphi\subseteq\mathbb{N}\times\mathbb{N}$
	\State $(\vect{L},\vect{A2})\,\gets\,\alnam{integrateStructuredPFD}(\vect{c},\vect{\gamma},\varphi)$
	// Note: $\vect{L}$ and $\vect{A2}$ are, respectively, vectors of logs and two-argument arctangent terms
	\State $\vect{A} \gets \alnam{atan2ToAtan}(\vect{A2})$ 
\EndIf
\State\Return $(P,g,\vect{L},\vect{A})$
\end{algorithmic}
\end{algorithm}

\s{Analysis of the Algorithm}
\label{algorithm-analysis}
We now consider the error analysis of the symbolic-numeric integration using LRT and PFD. We present a linear forward and backward error analysis for both methods.\footnote{Note that throughout this section we assume that the error for the numerical rootfinding for a polynomial $P(x)$ satisfies the relation $|\Delta r|\leq\varepsilon|r|$, where $r$ is the value of the computed root and $\Delta r$ is the distance in the complex plane to the exact root. This is accomplished using {\mpsolve} by specifying an error tolerance of $\varepsilon/\rm{deg}(P)$. Given the way that {\mpsolve} isolates roots, the bound is generally satisfied by several orders of magnitude.}

\bt[Backward Stability]
\label{backward-stability}
Given a rational function $f=A/B$ satisfying ${\rm deg}(A)<{\rm deg}(B)$, ${\rm gcd}(A,B)=1$ and input tolerance $\varepsilon$, Algorithm~\ref{symbolicNumericIntegrateLRT} and Algorithm~\ref{symbolicNumericIntegratePFD} yield an integral of a rational function $\hat{f}$ such that for $\Delta f=f-\hat{f}$, 
$$\|\Delta f\|_{\infty}=\emph{max}_x\left|\sum_k{\sf Re}\left(\Xi(x,r_k)\right)\right|+O(\varepsilon^2),$$
where the principal term is $O(\varepsilon)$, $r_k$ ranges over the evaluated roots and the function $\Xi$ defined below is computable. This expression for the backward error is finite on any closed, bounded interval not containing a root of $B(x)$.
\et

The advantage of exact computation on an approximate result is that the symbolic computation commutes with the approximation, \emph{i.e.}, we obtain the same result from issuing a given approximation and then computing symbolically as we do with computing symbolically first and then issuing the same approximation.\footnote{Although this comment is meant to explain the proof strategy, computing the symbolic result and then approximating also describes an alternative algorithm. Since this method requires lengthy computation of algebraic numbers that we would then approximate numerically anyway, we do not consider it.} Thus, we will conduct the error analysis throughout this section by assuming that we have exact information and then approximate at the end.

\bpf\ifThesis{\sf (PFD-based backward stability)}\else[PFD-based backward stability]\fi\vspace{12pt}
The PFD method begins by using Hermite reduction to obtain
\begin{equation}
\label{exact-integral}
\int f(x)\,dx=\frac{C(x)}{D(x)}+\int\frac{G(x)}{H(x)}\,dx,
\end{equation}
where $H(x)$ is squarefree. Given the roots $\gamma_i$ of $H(x)$ we may obtain the PFD of $G(x)/H(x)$, yielding
\begin{equation}
\label{pfd-expansion}
\frac{G(x)}{H(x)}=\sum_{i=1}^{\text{deg}(H)}\frac{c_i}{x-\gamma_i},
\end{equation}
where $c_i=c(\gamma_i)$ with $c(x)=G(x)/H'(x)$.  Taking account of identical residues, the expression \eqref{pfd-expansion} can then be integrated using the structured PFD algorithm described in Section \ref{review}. Since we approximate the roots of $H$, we replace the exact roots $\gamma_i$ with the approximations $\hat{\gamma}_i$. This breaks the symmetry of the exactly repeated residues, thus the (exact) $c_i$ are modified in two ways: by evaluating $c(x)$ at $\hat{\gamma}_i$; and restoring symmetry by adjusting the list of computed residues so that residues within $\varepsilon$ of each other are identified. This strategy requires some method of selecting a single representative for the list of nearby residues; the error analysis then estimates the error on the basis of the error of this representative.\footnote{Note that we assume that $\varepsilon$ is sufficiently small to avoid spurious identification of residues in this analysis. Even with spurious identification, however, the backward error analysis would only change slightly, \viz, to use the maximum error among the nearby residues, rather than the error of the selected representative residue.} We then represent this adjusted computed list of residues by $\hat{c}_i$. Since the Hermite reduction and PFD are equivalent to a rewriting of the input function $f(x)$ as
$$f(x)=\frac{C'(x)}{D(x)}-\frac{C(x)D'(x)}{D(x)^2}+\sum_{i=1}^{\text{deg}(H)}\frac{c_i}{x-\gamma_i},$$ 
the modified input $\hat{f}(x)$ that Algorithm~\ref{symbolicNumericIntegratePFD} integrates exactly is obtained from the above expression by replacing $c_i$ and $\gamma_i$ with $\hat{c}_i$ and $\hat{\gamma}_i$.

To compute the backward error we first must compute the sensitivity of the residues to changes in the roots. Letting $\Delta\gamma_i=\gamma_i-\hat{\gamma}_i$, then to first order we find that
$$c_i=c(\gamma_i)=c(\hat{\gamma}_i)+c'(\hat{\gamma}_i)\Delta\gamma_i+O(\Delta\gamma_i^2),$$
where $c'=\frac{G'}{H'}-\frac{GH''}{H'^2}$. 
So the backward error for a given term of the PFD is
\begin{eqnarray}
\frac{c_i}{x-\gamma_i}-\frac{\hat{c}_i}{x-\hat{\gamma}_i}&=&\frac{(c_i-\hat{c}_i)(x-\hat{\gamma}_i)+\hat{c}_i\Delta\gamma_i}{(x-\gamma_i)(x-\hat{\gamma}_i)}+O(\Delta\gamma_i^2)\\
&=&\frac{c'(\hat{\gamma}_i)\Delta\gamma_i}{(x-\hat{\gamma}_i-\Delta\gamma_i)}+\frac{\hat{c}_i\Delta\gamma_i}{(x-\hat{\gamma}_i)(x-\hat{\gamma}_i-\Delta\gamma_i)}+O(\Delta\gamma_i^2)\\
&=&\frac{c'(\hat{\gamma}_i)\Delta\gamma_i}{(x-\hat{\gamma}_i)}+\frac{\hat{c}_i\Delta\gamma_i}{(x-\hat{\gamma}_i)(x-\hat{\gamma}_i)}+O(\Delta\gamma_i^2).
\end{eqnarray}
Since any identified residues all approximate the same exact residue $c_k$, we use the error $c'(\gamma_k)$ for the residue $\hat{c}_k$ selected to represent the identical residues.

Now, because the rational part of the integral is computed exactly, only the PFD contributes to the backward error. Given that $\gamma_i$ is an exact root of $H(x)$
$$H(\gamma_i)=0=H(\hat{\gamma}_i)+H'(\hat{\gamma}_i)\Delta\gamma_i+O(\Delta\gamma_i^2),$$
where $H(\hat{\gamma}_i)\neq 0$ unless the exact root is computed, and $H'(\gamma_i)\neq 0$ (and hence $H'(\hat{\gamma}_i)\neq 0$) because $H$ is squarefree. Thus, we have that $\Delta\gamma_i=-H(\hat{\gamma}_i)/H'(\hat{\gamma}_i)$ to first order, where $|\Delta\gamma_i|\leq\varepsilon|\hat{\gamma}_i|$. We therefore find that
\begin{equation}
\Delta f=f-\hat{f}=
-\sum_{i=1}^{\text{deg}(H)}\left(\frac{c'(\hat{\gamma}_i)}{x-\hat{\gamma}_i}+\frac{\hat{c}_i}{(x-\hat{\gamma}_i)^2}\right)\frac{H(\hat{\gamma}_i)}{H'(\hat{\gamma}_i)}+O(\varepsilon^2).
\end{equation}
Since the summand is a rational function depending only on $x$ and $\hat{\gamma}_i$, for fixed $x$, the imaginary parts resulting from complex conjugate roots will cancel, so that only the real parts of the summand contribute to the backward error. We therefore find a first order expression of the backward error in the form of the theorem statement with
$$\boxed{\Xi(x,r_k)=\left(\frac{c'(r_k)}{x-r_k}+\frac{c(r_k)}{(x-r_k)^2}\right)\frac{H(r_k)}{H'(r_k)}},$$
which is $O(\varepsilon)$ because $\frac{H(r_k)}{H'(r_k)}$ is $O(\varepsilon)$. \ifLNCS\QEDB\fi
\epf
Note that, to properly account for the adjusted residue, applying the formula for $\Xi$ in the PFD case requires taking $r_k$ to be the $\gamma_k$ used to evaluate the representative residue.

\bpf\ifThesis{\sf (LRT-based backward stability) }\else[LRT-based backward stability]\fi 
The LRT algorithm produces an exact integral of the input rational function in the form
\begin{equation}
\label{exact-integral}
\int f(x)\,dx=\frac{C(x)}{D(x)}+\sum_{i=1}^n\sum_{c\, \mid\, U_i(t)\,=\,0}c\,\cdot\log(S_i(c,x)).
\end{equation}
Given a list $c_{ij}\in\mathbb{C}$, $1\leq j\leq\text{deg}(U_i)$ of roots of $U_i(t)$, we can express the integral in the form
$$\int f(x)\,dx=\frac{C(x)}{D(x)}+\sum_{i=1}^n\sum_{j=1}^{\text{deg}(U_i)}c_{ij}\,\cdot\log(S_i(c_{ij},x)),$$
where $n$ is the number of nontrivial squarefree factors of $\text{resultant}_x(H, G-cH')$.
Taking the derivative of this expression we obtain an equivalent expression of the input rational function as
\begin{equation}
\label{exact-equivalent-input-LRT}
f(x)=\frac{C'(x)}{D(x)}-\frac{C(x)D'(x)}{D(x)^2}+\sum_{i=1}^n\sum_{j=1}^{\text{deg}(U_i)}c_{ij}\frac{S_i'(c_{ij},x)}{S_i(c_{ij},x)}.
\end{equation}
The modified input $\hat{f}(x)$ that Algorithm~\ref{symbolicNumericIntegrateLRT} integrates exactly is obtained from this expression by replacing the exact roots $c_{ij}$ with their approximate counterparts~$\hat{c}_{ij}$.

To compute the backward error, we must compute the sensitivity of \eqref{exact-equivalent-input-LRT} to changes of the roots. Considering $f$ as a function of the parameters $c_{ij}$, and letting $\Delta c_{ij}=c_{ij}-\hat{c}_{ij}$, the difference between the exact root and the computed root, we find by taking partial derivatives with respect to the $c_{ij}$ that
\begin{multline}
f(x,c_{11},\ldots,c_{n\,\text{deg}(U_n)})=f(x,\hat{c}_{11},\ldots,\hat{c}_{n\,\text{deg}(U_n)}) +\\ 
\sum_{i=1}^n\sum_{j=1}^{\text{deg}(U_i)}\left.\left[\frac{\frac{\partial S_i(c,x)}{\partial x}}{S_i(c,x)}+c\left(\frac{\frac{\partial^2 S_i(c,x)}{\partial x\partial c}}{S_i(c,x)}-\frac{\frac{\partial S_i(c,x)}{\partial x}\frac{\partial S_i(c,x)}{\partial c}}{S_i(c,x)^2}\right)\right]\right|_{c=\hat{c}_{ij}}\Delta c_{ij}
+O(\Delta c_{ij}^2).
\end{multline}
Since $f(x,\hat{c}_{11},\ldots,\hat{c}_{n\,\text{deg}(U_n)})=\hat{f}(x)$, letting the rational function in square brackets be denoted by $\xi_i(c,x)$, we have that
$$\Delta f=f-\hat{f}=\sum_{i=1}^n\sum_{j=1}^{\text{deg}(U_i)}\xi_i(\hat{c}_{ij},x)\Delta c_{ij}
+O(\Delta\hat{c}_{ij}^2).$$
Given that $U_i(c_{ij})=0=U_i(\hat{c}_{ij})+U_i'(\hat{c}_{ij})\Delta c_{ij}+O(\Delta c_{ij}^2)$, we have that $\Delta c_{ij}=-U_i(\hat{c}_{ij})/U_i'(\hat{c}_{ij})$ to first order, where $|\Delta c_{ij}|\leq\varepsilon|\hat{c}_{ij}|$. Since, as for the PFD case, the imaginary terms from complex roots cancel, we therefore find a first order expression for the backward error in the form required by the theorem with
$$\boxed{\Xi(x,r_k)=\left.\left[\frac{\frac{\partial S_i(r,x)}{\partial x}}{S_i(r,x)}+r\left(\frac{\frac{\partial^2 S_i(r,x)}{\partial x\partial r}}{S_i(r,x)}-\frac{\frac{\partial S_i(r,x)}{\partial x}\frac{\partial S_i(r,x)}{\partial r}}{S_i(r,x)^2}\right)\right]\right|_{r=r_k}\frac{U_i(r_k)}{U_i'(r_k)}},$$
where $r_k$ runs over the roots $\hat{c}_{ij}$. This expression is $O(\varepsilon)$ because $\frac{U_i(r_k)}{U_i'(r_k)}$ is $O(\varepsilon)$. \ifLNCS\QEDB\fi
\epf

Note that the backward error is structured, because the manner in which the integral is computed preserves structure in the integrand for both the LRT-based Algorithm~\ref{symbolicNumericIntegrateLRT} and the PFD-based Algorithm~\ref{symbolicNumericIntegratePFD}. The use of Hermite reduction guarantees that the roots of the denominator of $\hat{f}(x)$ have the same multiplicity as the roots of $\hat{f}$. Then the identification of nearby computed residues in Algorithm~\ref{symbolicNumericIntegratePFD}, and the use of the Rothstein-Trager resultant in Algorithm~\ref{symbolicNumericIntegrateLRT}, ensures that the multiplicity of residues in the PFD of $G/H$ is also preserved, so that the PFD of $f$ and $\hat{f}$ have the same structure. This translates into higher degree arguments in the log and arctan terms of the integral than would be obtained by a standard PFD algorithm, leading to structured forward error as well.

It is important to reflect on the behaviour of these error terms $\Xi(x,r_k)$ near singularities of the integrand, which correspond to real roots of $H(x)$. For both algorithms, $\Xi$ contains a particular polynomial in the denominator that evaluates to zero at the real roots, specifically $x-\gamma_i$ and $S_i(c_{ij},x)$. In both cases, the expression of $\Xi$ has a term with the particular polynomial squared, which therefore asymptotically dominates the value of the error term near the singularity. This fact is important for efficient computation of the size of the error term near a singularity, since the scaling behaviour can be used to quickly locate the boundary around the singularity where the error starts to exceed the tolerance. Our implementation discussed in Section \ref{implementation} uses this scaling to compute such boundaries.

We turn now to the consideration of forward stability of the algorithms. We note that a full forward error analysis on this problem has subtleties on account of the numerical sensitivities of the log function. This does not affect the validity of the above analysis because near singularities the log term is dwarfed by the pole in the error term, so can be safely ignored in the computation of singularity boundaries. It is a concern when it comes to evaluation of the expressions of the integral.  This issue is reflected in the mastery that went into Kahan's ``atypically modest'' expression in \cite{kahan1980handheld}, which is written to optimize numerical stability of evaluation. We can, however, sidestep such concerns through the careful use of multiprecision numerics where the value is needed.

\bt[Forward Stability]
\label{forward-stability}
Given a rational function $f=A/B$ and tolerance $\varepsilon$, Algorithm~\ref{symbolicNumericIntegrateLRT} and Algorithm~\ref{symbolicNumericIntegratePFD} yield an integral of a rational function $\hat{f}$ in the form \eqref{tran-part} such that
$$\|\Delta{\textstyle\int} f\,dx\|_{\infty}=\emph{max}_x\left|\sum_k\left(\Xi(r_k,s_k,x)+\Theta(r_k,s_k,x)\right)\right|+O(\varepsilon^2),$$
where the leading term is $O(\varepsilon)$, $r_k$ and $s_k$ range over the real and imaginary parts of evaluated roots, and the functions $\Xi$ and $\Theta$ defined below, corresponding to log and arctangent terms, respectively, are computable. This expression for the forward error is finite on any closed, bounded interval not containing a root of $B(x)$.
\et
\bpf\ifThesis{\sf (LRT-based forward stability) }\else[LRT-based forward stability]\fi
We assume that we have computed the exact roots $c_{j\ell}$ of the $U_j(c)$ so that we can express the integral of the input rational function in the form
$$\int f(x)\,dx=\frac{C(x)}{D(x)}+\sum_{j=1}^n\sum_{\ell=1}^{\text{deg}(U_j)}c_{j\ell}\cdot\log(S_j(c_{j\ell},x)).$$
Since the roots $c_{j\ell}\in\mathbb{C}$, to get a real expression for the integral we can convert the transcendental part into a sum of logarithms and arctangents using the real and imaginary parts of the $c_{j\ell}$.

For the remainder of the proof we will assume that $c_k$ is a subsequence of the roots $c_{j\ell}$ of the squarefree factors of the Rothstein-Trager resultant such that each complex conjugate pair is only included once, and that $\varphi$ is a mapping defined by $k\mapsto j$ so that $S_{\varphi(k)}(c_k,x)$ is the term of the integral corresponding to the residue $c_k$. For each $c_k$ we let $a_k$ and $b_k$ be its real and imaginary parts, respectively. This allows us to express the integral in terms of logarithms and arctangent terms such that
\begin{equation}
\label{post-processed-integral}
\int f\,dx=\frac{C}{D}+\sum_{k=1}^m\left[a_k\log(V_k)+2b_k\,\text{arctan}(W_{1k},W_{2k})\right],\end{equation}
where $V_k$, $W_{1k}$ and $W_{2k}$ are functions of $a_k$, $b_k$ and $x$, and $m=\sum_{i=1}^n{\rm deg}(U_i)$.

Once again, since the rational part of the integral is computed exactly, it does not contribute to the forward error. The forward error is the result of the evaluation of the above expression at approximate values for the $a_k$ and $b_k$. Therefore, considering the variation of equation \eqref{post-processed-integral} with respect to changes in the $a_k$ and $b_k$ we obtain
\begin{multline}
\label{error-formula}
\Delta{\textstyle \int} f\,dx={\textstyle \int} (f-\hat{f})(x)\,dx=\\
\sum_{k=1}^m\left\{\left[\left(\frac{\partial V_k}{\partial a_k}\Delta a_k+\frac{\partial V_k}{\partial b_k}\Delta b_k\right)\frac{a_k}{V_k}+\log(V_k)\Delta a_k\right]+\right.\\ 
\left[\left(W_{2k}\frac{\partial W_{1k}}{\partial a_k}-W_{1k}\frac{\partial W_{2k}}{\partial a_k}\right)\Delta a_k+\right.\\
\left.\left(W_{2k}\frac{\partial W_{1k}}{\partial b_k}-W_{1k}\frac{\partial W_{2k}}{\partial b_k}\right)\Delta b_k\right]\frac{2b_k}{W_{1k}^2+W_{2k}^2}+\\
\left.2\text{arctan}(W_{1k},W_{2k})\Delta b_k\right\}+\text{h.o.t.}
\end{multline}

We now consider how to determine the values of $V_k$, $W_{1k}$, $W_{2k}$ and their partials from information in the computed integral. To simplify notation we let $j=\varphi(k)$. If $c_k$ is real, then we obtain a term of the form $a_k\log(S_j(a_k,x))$. In the complex case, each $c_k$ stands for a complex conjugate pair. As such, we obtain terms of the form
$$(a_k+i\,b_k)\log(S_j(a_k+i\,b_k,x))+(a_k-i\,b_k)\log(S_j(a_k-i\,b_k,x).$$
Expressing $S_j(a_k+i\,b_k,x)$ in terms of real and imaginary parts as $W_{1k}(x)+i\,W_{2k}(x)\equiv W_{1k}(a_k,b_k,x)+i\,W_{2k}(a_k,b_k,x)$, so that $S_j(a_k-i\,b_k,x)=W_{1k}(x)-i\,W_{2k}(x)$, the expression of the term in the integral becomes
or
$$a_{k}\log\left(W_{1k}(x)^2+W_{2k}(x)^2\right)+i\,b_{k}\log\left(\frac{W_{1k}(x)+i\,W_{2k}(x)}{W_{1k}(x)-i\,W_{2k}(x)}\right).$$
The observation that $i\log\left(\frac{X+iY}{X-iY}\right)$ has the same derivative as $2\,\text{arctan}(X,Y)$ 
 allows the term of the integral to be converted into the form of the summand in \eqref{post-processed-integral} with
$V_k(x) =W_{1k}(x)^2+W_{2k}(x)^2$.

We can express $V_k$, $W_{1k}$ and $W_{2k}$ and their partials in terms of $S_j(c,x)$ and $\partial S_j(c,x)/\partial c$ as follows. First of all we have that
\begin{equation}
\label{fea-parts}
W_{1k}(x)=\mathsf{Re}(S_j(c_k,x)), \quad W_{2k}(x)=\mathsf{Im}(S_j(c_k,x)).
\end{equation} 
Then, because $c$ is an indeterminate in $S_j(c,x)$, 
$\left.\frac{\partial S_j(c,x)}{\partial c}\right|_{c=c_k}=\frac{\partial S_j(c_k,x)}{\partial a_k}$ with $\frac{\partial S_j(c_k,x)}{\partial a_k}=\frac{\partial W_{1k}(c_k,x)}{\partial a_k}+i\,\frac{\partial W_{2k}(c_k,x)}{\partial a_k},$ so that  
\begin{equation}
\label{fea-partials}
\frac{\partial W_{1k}}{\partial a_k}=\textsf{Re}\left(\left.\frac{\partial S_j(c,x)}{\partial c}\right|_{c=c_k}\right),\quad\frac{\partial W_{2k}}{\partial a_k}=\textsf{Im}\left(\left.\frac{\partial S_j(c,x)}{\partial c}\right|_{c=c_k}\right).
\end{equation}
In a similar way, and because the derivative w.r.t. $b_k$ picks up a factor of $i$, $\frac{\partial W_{1k}}{\partial b_k}=-\frac{\partial W_{2k}}{\partial a_k}$ and $\frac{\partial W_{2k}}{\partial b_k}=\frac{\partial W_{1k}}{\partial a_k}$. It follows, then, that
\ifThesis
$$\frac{\partial V_k}{\partial a_k} = 2\left(W_{1k}\frac{\partial W_{1k}}{\partial a_k}+W_{2k}\frac{\partial W_{2k}}{\partial a_k}\right)\ \text{ and }\ \frac{\partial V_k}{\partial b_k} = 2\left(W_{2k}\frac{\partial W_{1k}}{\partial a_k}-W_{1k}\frac{\partial W_{2k}}{\partial a_k}\right).$$
\else
$\frac{\partial V_k}{\partial a_k} = 2\left(W_{1k}\frac{\partial W_{1k}}{\partial a_k}+W_{2k}\frac{\partial W_{2k}}{\partial a_k}\right)$ and $\frac{\partial V_k}{\partial b_k} = 2\left(W_{2k}\frac{\partial W_{1k}}{\partial a_k}-W_{1k}\frac{\partial W_{2k}}{\partial a_k}\right)$. 
\fi

For the complex root case, given the error bound $|\Delta c|\leq \varepsilon |\hat{c}|$ on the complex roots, we have the same bound on the real and imaginary parts, \emph{viz.}, $|\Delta a|\leq \varepsilon |\hat{a}|$, $|\Delta b|\leq \varepsilon |\hat{b}|$. Since $\Delta c_{k}=-U_j(\hat{c}_k)/U_j'(\hat{c}_k)$ to first order, and $\Delta c_k=\Delta a_k+i\Delta b_k$, from \eqref{error-formula} we therefore obtain an expression for the linear forward error in the form required by the theorem with
$$\boxed{\Xi(\hat{a}_k,\hat{b}_k,x) = \left(2a_k\Gamma+\log\left(W_{1k}^2+W_{2k}^2\right)\right)\textsf{Re}\left(\frac{U_j}{U_j'}\right)+2a_k\Lambda\,\textsf{Im}\left(\frac{U_j}{U_j'}\right)}$$
when $b_k\neq 0$, otherwise $\Xi(\hat{a}_k,\hat{b}_k,x)\equiv 0$, and with
$$\boxed{\Theta(\hat{a}_k,\hat{b}_k,x) =2b_k\Lambda\,\textsf{Re}\left(\frac{U_j}{U_j'}\right)+2\left(\textrm{artcan}\left(W_{1k},W_{2k}\right)-b_k\Gamma\right)\textsf{Im}\left(\frac{U_j}{U_j'}\right)},$$
where $\Gamma=\frac{W_{1k}\frac{\partial W_{1k}}{\partial a_k}+W_{2k}\frac{\partial W_{2k}}{\partial a_k}}{W_{1k}^2+W_{2k}^2}$, $\Lambda=\frac{W_{2k}\frac{\partial W_{1k}}{\partial a_k}-W_{1k}\frac{\partial W_{2k}}{\partial a_k}}{W_{1k}^2+W_{2k}^2}$, $W_{1k}$ and $W_{2k}$ are given by \eqref{fea-parts}, $\frac{\partial W_{1k}}{\partial a_k}$ and $\frac{\partial W_{2k}}{\partial a_k}$ are given by \eqref{fea-partials}, and $U_j$ and $U'_j$ are evaluated at $\hat{c}_k=\hat{a}_k+i\, \hat{b}_k$. These terms are $O(\varepsilon)$ because $\frac{U_j(\hat{c}_k)}{U'_j(\hat{c}_k)}$ is $O(\varepsilon)$.

For the real root case we have a much simpler expression, since $\Theta(\hat{a}_k,\hat{b}_k,x)\equiv 0$ and since $\hat{c}_k = \hat{a}_k$,
$$\boxed{\Xi(\hat{a}_k,\hat{b}_k,x) =\left(\hat{a}_k\frac{\left.\frac{\partial S_j}{\partial c}\right|_{c=\hat{a}_k}}{S_j(\hat{a}_k,x)}+\log(S_j(\hat{a}_k,x))\right)\frac{U_j(\hat{\alpha}_k)}{U_j'(\hat{\alpha}_k)}},$$
which is also $O(\varepsilon)$. \ifLNCS\QEDB\fi
\epf

\bpf\ifThesis{\sf (PFD-based forward stability) }\else[PFD-based forward stability]\fi
Proceeding as we did for the LRT method, if we assume that the roots of the denominator of the polynomial $H(x)$ are computed exactly, then we obtain an exact expression of the integral of $f$ in the form
\begin{equation}
\int f(x)\,dx=\frac{C(x)}{D(x)}+\sum_{i=1}^{\text{deg}(H)}c_i(\gamma_i)\log(x-\gamma_i).
\end{equation}
As in the LRT-based proof, we assume $\gamma_k$ is a subsequence of the $\gamma_i$ that includes only one conjugate of each complex root. Then the same techniques for converting this to a sum of logarithms and arctangents can be applied here. Since $H(x)$ is squarefree, all of the $\gamma_k=\alpha_k+i\,\beta_k$ are simple roots, which entails that the integral can be expressed in the form \eqref{post-processed-integral} where the $V_j(x)$ are equal to $x-\alpha_k$ for a real root and $x^2-2\alpha_k+\alpha_k^2+\beta_k^2$ for a complex root with $a_k=\mathsf{Re}\left(c(\gamma_k)\right)$, with $c(x)=G(x)/H'(x)$, and using Rioboo's trick the $W_{1k}(x)=\alpha_k-x$ and $W_{2k}=\beta_k$ and $b_k=\mathsf{Im}\left(c(\gamma_k)\right)$. Even though the structured integral is not expressed in this form, it is still an exact integral that we approximate, where all subsequent computation we perform is exact. Analyzing the error in this form has the advantage of using information available after the completion of the rootfinding task. Thus, we will analyze the forward error in this form.

Because the residues are now obtained by computation, and we find the roots of $H(x)$, we obtain a modified version of the first order forward error formula \eqref{error-formula}, \emph{viz.},
\begin{multline}
\label{error-formula2}
\Delta{\textstyle \int} f\,dx={\textstyle \int} (f-\hat{f})(x)\,dx=\\
\sum_{k=1}^m\left\{\left[\left(\frac{\partial V_k}{\partial \alpha_k}\Delta \alpha_k+\frac{\partial V_k}{\partial \beta_k}\Delta \beta_k\right)\frac{a_k}{V_k}+\left(\frac{\partial a_k}{\partial \alpha_k}\Delta \alpha_k+\frac{\partial a_k}{\partial \beta_k}\Delta \beta_k\right)\log(V_k)\right]+\right.\\ \left.\frac{2\beta_kb_k(\Delta\alpha_k+\Delta\beta_k)}{(\alpha_k-x)^2+\beta_k^2}+2\left(\frac{\partial b_k}{\partial \alpha_k}\Delta \alpha_k+\frac{\partial b_k}{\partial \beta_k}\Delta \beta_k\right)\text{arctan}(\alpha_k-x,\beta_k)\right\}+\text{h.o.t.}
\end{multline}

Since $c(x)=G(x)/H'(x)$, $c'(x)=\frac{G'(x)}{H'(x)}-\frac{G(x)H''(x)}{H'(x)^2}$, and so it follows that
$\frac{\partial a_k}{\partial \alpha_k}=\mathsf{Re}(c'(\gamma_k))$ and $\frac{\partial b_k}{\partial \alpha_k}=\mathsf{Im}(c'(\gamma_k))$. Similarly, $\frac{\partial a_k}{\partial \beta_k}=-\mathsf{Im}(c'(\gamma_k))$ and $\frac{\partial b_k}{\partial \beta_k}=\mathsf{Re}(c'(\gamma_k))$. For the complex root  case, then, since $\Delta\gamma_j=-H(\hat{\gamma}_j)/H'(\hat{\gamma}_j)$ to first order, we obtain from equation \eqref{error-formula2} an expression for the linear forward error in the form required by the theorem  with $\Xi(\hat{\alpha}_k,\hat{\beta}_k,x)=\Xi_a+\Xi_b$, where
$$\boxed{\Xi_a=\left(\frac{2\hat{a}_k(\hat{\alpha}_k-1)}{(\hat{\alpha}_k-x)^2+\hat{\beta}_k^2}+ \mathsf{Re}\left(c'(\hat{\gamma}_k)\right)\log\left((\hat{\alpha}_k-x)^2+\hat{\beta}_k^2\right)\right)\mathsf{Re}\left(\frac{H}{H'}\right)}$$
and
$$\boxed{\Xi_b=\left(\frac{2\hat{a}_k\hat{\beta}_k}{(\hat{\alpha}_k-x)^2+\hat{\beta}_k^2}+ \mathsf{Im}\left(c'(\hat{\gamma}_k)\right)\log\left((\hat{\alpha}_k-x)^2+\hat{\beta}_k^2\right)\right)\mathsf{Im}\left(\frac{H}{H'}\right)}$$
when $b_k\neq 0$, otherwise $\Xi(\hat{\alpha}_k,\hat{\beta}_k,x)\equiv 0$, and with $\Theta(\hat{\alpha}_k,\hat{\beta_k},x)=\Theta_a+\Theta_b$, where
$$\boxed{\Theta_a=\left(\frac{2\hat{\beta_k}\hat{b}_k}{(\hat{\alpha}_k-x)^2+\hat{\beta}_k^2}- \mathsf{Im}\left(c'(\hat{\gamma}_k)\right)\text{arctan}(\hat{\alpha}_k-x,\hat{\beta}_k)\right)\mathsf{Re}\left(\frac{H}{H'}\right)}$$
and
$$\boxed{\Theta_b=\left(\frac{2\hat{\beta_k}\hat{b}_k}{(\hat{\alpha}_k-x)^2+\hat{\beta}_k^2}+ \mathsf{Re}\left(c'(\hat{\gamma}_k)\right)\text{arctan}(\hat{\alpha}_k-x,\hat{\beta}_k)\right)\mathsf{Im}\left(\frac{H}{H'}\right)},$$
with $H$ and $H'$ being evaluated in all cases at $\hat{\gamma}_k$. All of these terms are $O(\varepsilon)$ because $\frac{H}{H'}$ is.

In the case of real roots,
$$\boxed{\Xi(\hat{\alpha}_k,\hat{\beta}_k,x)=\left( c'(\hat{\alpha}_k)\log\left(x-\hat{\alpha}_k\right)-\frac{\hat{a}_k}{x-\hat{\alpha}_k}\right)\frac{H(\hat{\alpha}_k)}{H'(\hat{\alpha}_k)}},$$
which is also $O(\varepsilon)$. \ifLNCS\QEDB\fi

\epf

We note again that the forward error is structured for both algorithms. In the LRT-based case, the exact integral is computed and the approximation only perturbs the values of coefficients of polynomials in the integral, with all symmetries in the computed integral being preserved. In the PFD-based case this comes out in the identification of the residues that are with $\varepsilon$ of each other. This means that for whichever $\tilde{k}$ is chosen for the representative residue, then $\hat{a}_{\tilde{k}}$, $\hat{b}_{\tilde{k}}$, and $c'(\gamma_{\tilde{k}})$ must be used to evaluate the error terms corresponding to each of the roots that have the same residue.

Once again, note that the scaling behaviour for the error term for real roots can be used to efficiently compute the boundaries around the singularities in the integral. In this case, the error scales as $(x-\alpha)^{-1}$ and $S_j(a_k,x)^{-1}$, since the quadratic terms appearing the backward error have been integrated. As a result, the forward error grows much more slowly as we approach a singularity and we get much smaller bounds before the error exceed the tolerance.
\s{Implementation}
\label{implementation}

We have implemented the algorithms presented in Section~\ref{algorithm}.  
In our code, the symbolic computations are realized with the {\em Basic
Polynomial Algebra Subprograms} (\textsc{bpas})\label{symbol:bpas} publicly available in
source at \url{http://bpaslib.org}.  The \textsc{bpas} library offers
polynomial arithmetic operations (multiplication, division, root
isolation, etc.) for univariate and multivariate polynomials with
integer, rational or complex rational number coefficients; it is
written in C++ with CilkPlus extension for optimization on multicore
architectures and built on top of the GMP library.

The numerical portion of our computation relies on \textsc{MPSolve},\label{symbol:mpsolve}
publicly available in source
at \url{http://numpi.dm.unipi.it/mpsolve}.  The \textsc{MPSolve}
library, which is written in C and built upon the GMP library, offers
arbitrary precision solvers for polynomials and secular equations,
\emph{a posteriori} 
guaranteed inclusion radii, even on restricted domains; for requested
output precision of $2^{-w}$, it provides $w$ correct digits in the
returned roots (see \cite{bini2014solving} for more details).

The implementation of both Algorithm~\ref{symbolicNumericIntegrateLRT} and Algorithm~\ref{symbolicNumericIntegratePFD} are integrated into the
\textsc{bpas} library; Algorithms~\ref{symbolicNumericIntegrateLRT} and \ref{symbolicNumericIntegratePFD} can be called, respectively, through the
{\tt real\-Sym\-bo\-lic\-Nu\-me\-ric\-In\-te\-grate} and {\tt real\-Sym\-bo\-lic\-Nu\-me\-ric\-In\-te\-grate\-PFD}
methods of the {\tt Uni\-va\-ria\-te\-Ra\-tio\-nal\-Function} template class. We abbreviate the {\tt real\-Sym\-bo\-lic\-Nu\-me\-ric\-In\-te\-grate} method to {\tt snIntLRT}\label{symbol:snintlrt} and the {\tt real\-Sym\-bo\-lic\-Nu\-me\-ric\-In\-te\-grate\-PFD} method as {\tt snIntPFD}\label{symbol:snintpfd} in the sequel. 
The following output formats are available in the {\tt Uni\-va\-ria\-te\-Ra\-tio\-nal\-Function} class:
approximate (floating point number)
or symbolic (rational number) expressions (in either \textsc{Maple} or \textsc{Matlab}
syntax); see Figure~\ref{sample-output} for
a combination of floating point and \textsc{Maple} output formats.

\bfig[ht!]
\vspace{-12pt}
\bc
\ifThesis
\includegraphics[scale=0.28]{\paperpath/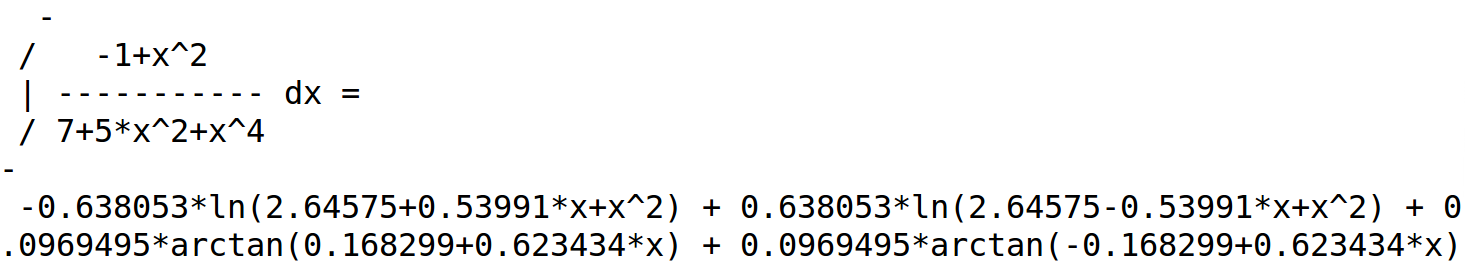}
\else
\includegraphics[scale=0.228]{\paperpath/BPAS-integral.png}
\fi
\vspace{-18pt}
\ec
\caption{Sample output of {\tt snInt}.}
\label{sample-output}
\efig

For the integral
\ifLNCS
$\int\frac{(x^2-1)dx}{x^4+5x^2+7},$
\fi
\textsc{Maple} provides the expression appearing in Figure \ref{maple-integral}. For the same integral, the \textsc{bpas}/\textsc{MPSolve} routines \texttt{snIntLRT} and \texttt{snIntPFD} both return the output shown in Figure~\ref{sample-output} in the default floating point output format. In the data structures, however, the coefficients are stored as multiprecision rational numbers, which can be displayed by changing the output format.

It must be noted that there are differences between Algorithms~\ref{symbolicNumericIntegrateLRT} and \ref{symbolicNumericIntegratePFD} and their implementations in \bpas. The key difference is that {\tt snIntPFD} and {\tt snIntLRT} do additional post-processing. As such, the forward error analysis detailed in section \ref{algorithm-analysis} assumes a different output expression than is produced finally in the implementations. Both {\tt snIntPFD} and {\tt snIntLRT} do compute the integral in the form assumed in the forward error analysis, however. There are therefore several reasons why the additional post-processing will not significantly affect the conclusions drawn from the error analysis.

First of all, after the integral is computed in the form of equation 
\eqref{post-processed-integral}, all further computation in \bpas\ is done using exact computation. As such, the final expression, which uses Rioboo's method to remove spurious singularities from the arctangents, is mathematically equivalent to the integral in the form of \eqref{post-processed-integral} from the perspective of the integration problem, \viz, they have the same derivative and hence differ only by a constant.

Another reason why the additional post-processing will not affect the forward error evaluation is that converting two-argument arctangent functions of polynomials (or one-argument arctangents of rational functions) to one-arguments arctangents of polynomials increases their numerical stability. This is because the derivative of $\arctan(x)$ is $1/(1+x^2)$, which can never be zero, or even small, for real integrals, whereas the derivative of $\arctan(x_1,x_2)$ and $\arctan(x_1/x_2)$ is $(x_1'x_2-x_1x_2')/(x_1^2+x_2^2)$, which can approach zero for nearly real roots of the denominator of the integrand. This changes the denominators of the expressions for $\Theta(r_k,s_k,x)$ appearing in the proof of theorem \ref{forward-stability}. Thus, the application of Rioboo's method improves the stability of the integral. As such, the worst that can happen in this situation is that the forward error looks to become large when it is not. Though this issue will need to be resolved in refinements of the implementation, it is very unlikely to be a significant issue on account of the fact that the error is dominated by the error in the roots, and in practice this error is many orders of magnitude less than the tolerance.

Since the forward error analysis is reliable, modulo the issue just stated, even though we could compute the forward error on the final output, it is a design decision not to do so. This is because a design goal of the algorithm is to hide the error analysis in the main computation of the integral by performing the error analysis in parallel. This is only possible if the error analysis can proceed on information available before the integral computation is completed.

\s{Experimentation}
\label{experiments}

We now consider the performance of Algorithms~\ref{symbolicNumericIntegrateLRT} and \ref{symbolicNumericIntegratePFD} based on their implementations in \bpas.\footnote{The data for this section was collected in December of 2017.} For the purposes of comparing their runtime performance we will consider integration of the following functions:
\be
\item $f_1(x)=\frac{1}{x^n-2}$;\ifThesis\vspace{-6pt}\else\fi
\item $f_2(x)=\frac{1}{x^n+x-2}$;\ifThesis\vspace{-6pt}\else\fi
\item $f_3(x)=[n,n]_{e^x/x}(x)$,
\ee
where $[m,n]_f(x)$\label{symbol:pade} denotes the Pad\'{e} approximant of order $[m/n]$ of $f$. Since
$\int \frac{e^x}{x}dx=\text{Ei}(x),$
the non-elementary exponential integral, integrating $f_3$ provides a way of approximating $\text{Ei}(x)$. These three problems test different features of the integrator on account of the fact that $f_1(x)$ has a high degree of symmetry, while $f_2(x)$ breaks this symmetry, and $f_3(x)$ contains very large integer coefficients for moderate size $n$. Note that unless otherwise stated, we are running  {\tt snIntPFD} and {\tt snIntLRT} with the error analysis computation turned on.

Comparing {\tt snIntPFD} and {\tt snIntLRT} on functions $f_1$ and $f_2$ for Fibonacci values of $n$, from $n=3$ to $n=377$, we find the results shown in Figure \ref{fig:runtime-comp}. We see from Figure \ref{fig:runtime-comp-f1} that the performance of the two algorithms is nearly identical on function $f_1(x)$. Figure \ref{fig:runtime-comp-f2} shows, however, that on function $f_2(x)$, {\tt snIntPFD} performs considerably poorer than {\tt snIntLRT}. The reason for this is that the size of the coefficients in the Rothstein-Trager resultant grows exponentially for $f_2(x)$. This causes no significant issues for the subresultant computation, but it significantly slows the rootfinding algorithm, leading to large rational number roots, which slows the post-processing algorithms. In contrast, the difference in runtime for {\tt snIntPFD} on functions $f_1(x)$ and $f_2(x)$ is negligible. This is because the speed of {\tt snIntPFD} is determined by the degree of the denominator (after the squarefree part has been extracted by Hermite reduction) and the height of the coefficients. Since the denominators of $f_1$ and $f_2$ have the same degree and height bound, we expect the performance to be similar.

\bfig[h!]
\centering
\ifThesis
\begin{subfigure}[b]{0.82\textwidth}
\else
\begin{subfigure}[b]{0.88\textwidth}
\fi
\centering
\includegraphics[width=\textwidth]{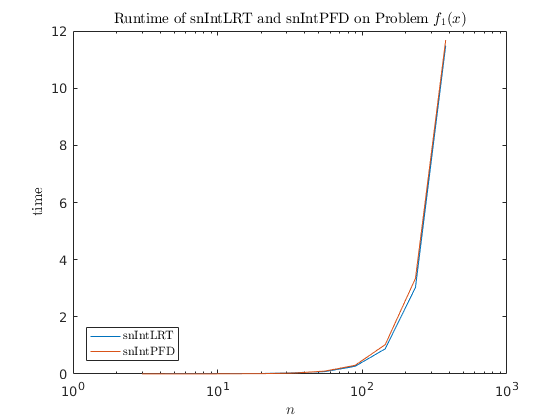}
\caption{}
\label{fig:runtime-comp-f1}
\end{subfigure}
\ifThesis
\begin{subfigure}[b]{0.82\textwidth}
\else
\begin{subfigure}[b]{0.88\textwidth}
\fi
\centering
\includegraphics[width=\textwidth]{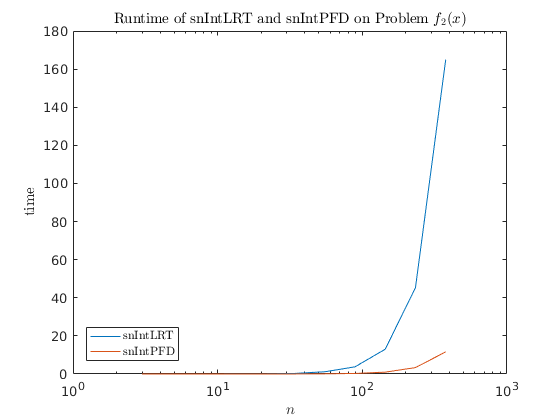}
\caption{}
\label{fig:runtime-comp-f2}
\end{subfigure}
\caption[Runtime comparison of {\tt snIntPFD} and {\tt snIntLRT} for integrals of $\frac{1}{x^n-2}$ and $\frac{1}{x^n+x-2}$]{Runtime comparison of {\tt snIntPFD} and {\tt snIntLRT} on problems (a) $f_1(x)$ and (b) $f_2(x)$.}
\label{fig:runtime-comp}
\efig

If we run the same computation with the error analysis turned off, we see that {\tt snIntLRT} actually performs better than {\tt snIntPFD} on problem $1$. With the performance improvement of {\tt snIntLRT} relative to {\tt snIntPFD} being similar to the performance of {\tt snIntPFD} relative to {\tt snIntLRT} on problem 2 with the error analysis turned on. Thus, there are some problems on which {\tt snIntLRT} performs better than {\tt snIntPFD}. The performance of {\tt snIntPFD} is easier to predict from the degree and height bound of the input polynomials.

That there is a difference in performance in  {\tt snIntLRT} when the error analysis computation is turned off shows that the current implementation can only partially hide the error analysis computation for {\tt snIntLRT} on some problems. The error analysis computation is successfully completely hidden for problem 2 with  {\tt snIntLRT}, which is to be expected. For  {\tt snIntPFD}, however, there is a negligible difference in the runtime with the error analysis turned on and off. Thus, once again, {\tt snIntPFD} has the more reliable and desirable behaviour.

{\tt snIntPFD} also performs better on problem $3$, which leads to coefficients with height bound that grows exponentially with $n$. For $n=8$, {\tt snIntLRT} computes the integral in about $0.04$ s, whereas  {\tt snIntPFD} computes it in about $0.01$ s. For $n=13$, the respective runtimes increase to $0.18$ s and $0.02$ s, and by $n=21$, around $2.5$ s and $0.04$ s. This shows that {\tt snIntLRT} is considerably slowed down by large input coefficients, since this leads to even larger coefficients in the subresultants. This is reflected in the subresultant computation taking $0.6$ s for $n=21$ and slowing down the exact integration to $2.4$ s. Thus, when it comes to runtime performance,  {\tt snIntPFD} is the clear winner.

Turning to the error analysis, we now consider the behaviour of the error under variation of the input tolerance $\varepsilon$. For integrands without real singularities, we can compute a global forward and backward error bound over the entire real line. For the non-singular problem $\int \frac{dx}{x^{128}+2}$, a variant of problem 1, we see from table \ref{tolerance-proportionality-regular} that both {\tt snIntLRT} and {\tt snIntPFD} exhibit tolerance proportionality as the tolerance is reduced. Here {\tt snIntLRT} generally outperforms {\tt snIntPFD} for a given input tolerance by several orders of magnitude, but both algorithms perform strongly.
\begin{table}
\centering
\begin{tabular}{@{}lcccc@{}}\toprule
&\multicolumn{2}{c}{\tt snIntLRT}&\multicolumn{2}{c}{\tt snIntPFD}\\ \midrule
\multicolumn{1}{c}{$\varepsilon$}&\mbox{ forward error }& \mbox{ backward error }&\mbox{ forward error }&\mbox{ backward error }\\
\midrule
$6\cdot 10^{-11}$ $(2^{-34})$&$8\cdot 10^{-16}$&$1\cdot 10^{-15}$&$2\cdot 10^{-15}$&$2\cdot 10^{-12}$\\
$3\cdot 10^{-17}$ $(2^{-55})$&$3\cdot 10^{-55}$&$2\cdot 10^{-53}$&$1\cdot 10^{-39}$&$1\cdot 10^{-38}$\\
$2\cdot 10^{-27}$ $(2^{-89})$&$1\cdot 10^{-75}$&$2\cdot 10^{-73}$&$9\cdot 10^{-59}$&$8\cdot 10^{-58}$\\
$4\cdot 10^{-44}$ $(2^{-144})$&$6\cdot 10^{-95}$&$7\cdot 10^{-93}$&$3\cdot 10^{-77}$&$2\cdot 10^{-76}$\\
\bottomrule
\end{tabular}
\vspace{4pt}
\caption{Tolerance proportionality of the global forward and backward error bounds for {\tt snIntLRT} and {\tt snIntPFD} on $\int \frac{dx}{x^{128}+2}$.}
\label{tolerance-proportionality-regular}
\end{table}

On problems that do have real singularities, we obtain boundaries around the singularities past which the error exceeds the input tolerance. On problem 3 for $n=8$, there is a real singularity at $x\stackrel{\cdot}{=}10.949$. For this singularity, we see from table \ref{tolerance-proportionality-singular} that both {\tt snIntLRT} and {\tt snIntPFD} exhibit tolerance proportionality of the singularity boundaries as the tolerance is reduced. Thus, we can get as close to the singularity as desired by decreasing the input tolerance. With the exception of $\varepsilon=2^{-34}$, {\tt snIntLRT} outperforms {\tt snIntPFD}, but the difference in performance between the two algorithms is not as extreme as with the non-singular case. For the default precision of $\varepsilon=2^{-53}$ and above, both algorithms get extremely close to the singularity before the error exceeds the tolerance.

\begin{table}
\centering
\begin{tabular}{@{}lcccc@{}}\toprule
&\multicolumn{2}{c}{\tt snIntLRT}&\multicolumn{2}{c}{\tt snIntPFD}\\ \midrule
\multicolumn{1}{c}{$\varepsilon$}&\mbox{ forward error $\partial$ }& \mbox{ backward error $\partial$ }&\mbox{ forward error $\partial$ }&\mbox{ backward error $\partial$ }\\
\midrule
$2^{-34}$&$4\cdot 10^{-3}$&$6\cdot 10^{-2}$&$1\cdot 10^{-14}$&$6\cdot 10^{-7}$\\
$2^{-55}$&$7\cdot 10^{-23}$&$9\cdot 10^{-12}$&$8\cdot 10^{-20}$&$3\cdot 10^{-10}$\\
$2^{-89}$&$4\cdot 10^{-32}$&$2\cdot 10^{-16}$&$2\cdot 10^{-28}$&$1\cdot 10^{-14}$\\
$2^{-144}$&$2\cdot 10^{-34}$&$1\cdot 10^{-17}$&$3\cdot 10^{-31}$&$6\cdot 10^{-16}$\\
\bottomrule
\end{tabular}
\vspace{4pt}
\caption[Tolerance proportionality of the singularity boundary widths for \texttt{snIntLRT} and \texttt{snIntPFD} on $\int\mbox{[8,8]}_{e^x/x}(x)dx$ for the singularity at $x\stackrel{\cdot}{=}10.949$.]{Tolerance proportionality of the singularity boundary widths for {\tt snIntLRT} and {\tt snIntPFD} on problem 3 with $n=8$ for the singularity at $x\stackrel{\cdot}{=}10.949$. The symbol $\partial$ is used to abbreviate ``boundary width''.}
\label{tolerance-proportionality-singular}
\end{table}

For testing the numerical stability, we will consider two additional problems, along with problem 3 above:
\be
\item[4.] $f(x)=\frac{2x}{x^2-(1+\epsilon)^2}, \epsilon\rightarrow 0$ (singular just outside $[-1,1]$);;\ifThesis\vspace{-6pt}\else\fi
\item[5.] $f(x)=\frac{2x}{x^2+\epsilon^2}, \epsilon\rightarrow 0$ (nearly real singularities on the imaginary axis).
\ee
Note that the small parameter $\epsilon$ in problems 4 and 5 is conceptually distinct from the input tolerance $\varepsilon$. These problems are useful for testing the stability of the integration algorithms near singularities.

On problems $4$ and $5$, {\tt snIntLRT} computes the exact integral, because the integral contains only rational numbers, so there is no need to do any rootfinding. Thus, the forward and backward error are exactly zero, and the evaluation of the integral is insensitive to how close the singularities are to the boundary of the interval of integration, provided a numerically stable method of evaluating the logarithm is used. 

On the same problems {\tt snIntPFD} computes very nearly the exact integral. On problem 4, with $\varepsilon=2^{-53}$, it is possible to get to within about $1.6\cdot10^{-23}$ of the singularities at $\pm1\pm\epsilon$ before the error exceeds the tolerance. Thus, even with $\epsilon=\varepsilon$, the error does not affect the evaluation of the integral on the interval $[-1,1]$. {\tt snIntPFD} also performs exceedingly well on problem 5. With the same input tolerance, the forward error bound is $1.9\cdot 10^{-57}$ for $\epsilon=0.1$ and increases only to $1.7\cdot10^{-42}$ for $\epsilon=10^{-16}$. Indeed, the difference between the $a$ in $\log(x^2+a)$ computed by {\tt snIntLRT} and {\tt snIntPFD} is about $1.7\cdot 10^{-74}$.

Since problem 3 requires rootfinding for both algorithms, it provides a more fair comparison. {\tt snIntLRT} fares slightly better than {\tt snIntPFD} on this problem, but only slightly and not in a way that significantly affects numerical evaluation. We make the comparison for $\varepsilon=2^{-53}$. With $n=8$, for {\tt snIntLRT} the backward and forward error bounds away from the real singularities are about $2.1\cdot 10^{-38}$ and $1.8\cdot 10^{-36}$, respectively. For {\tt snIntPFD} the backward error bound increases only to $2.6\cdot 10^{-35}$ and the forward error bound decreases slightly to $1.2\cdot 10^{-36}$. As for evaluation near the real singularities, {\tt snIntLRT} can get within about $1.8\cdot 10^{-23}$ of the singularity at around $x=10.949$ before the forward error exceeds the tolerance. {\tt snIntPFD} can get within about $2\cdot 10^{-20}$. Of course, this is not a true concern anyway, because the Pad\'{e} approximant ceases to be a good approximation of $e^x/x$ before reaching the real root.

We see, therefore, that {\tt snIntPFD} performs strongly against {\tt snIntLRT} even when {\tt snIntLRT} gets the exact answer and {\tt snIntPFD} does not. Indeed, the differences in numerical stability between the two methods are relatively small. Given the performance benefits of {\tt snIntPFD}, the PFD-based algorithm is the clear overall winner between the two algorithms. {\tt snIntPFD} is therefore the preferred choice, except in cases where the exact integral needs to be retained.

\s{Conclusion}
We have identified two methods for the hybrid symbolic-numeric integration of rational functions on exact input that adjust the forward and backward error the integration according to a user-specified tolerance, determining the intervals of integration on which the integration is numerically stable. The PFD-based method is overall the better algorithm, being better overall in terms of runtime performance while maintaining excellent numerical stability. The LRT-based method is still advantagous in contexts where the exact integral needs to be retained for further symbolic computation. We believe these algorithms, and the extension of this approach to wider classes of integrands, has potential to increase the utility of symbolic computation in scientific computing.

\bibliography{symbint}
\bibliographystyle{plain}

\end{document}